\documentstyle[preprint,tighten,pra,aps]{revtex}
\begin{document} \draft

\topmargin -.6in
\def\br{\begin{eqnarray}}
\def\er{\end{eqnarray}}
\def\be{\begin{equation}}
\def\ee{\end{equation}}
\def\({\left(}
\def\){\right)}
\def\a{\alpha}
\def\b{\beta}
\def\d{\delta}
\def\D{\Delta}
\def\g{\gamma}
\def\G{\Gamma}
\def\h{ {1\over 2}  }
\def\hp{ {+{1\over 2}}  }
\def\hm{ {-{1\over 2}}  }
\def\k{\kappa}
\def\l{\lambda}
\def\L{\Lambda}
\def\m{\mu}
\def\n{\nu}
\def\o{\over}
\def\O{\Omega}
\def\p{\phi}
\def\rh{\rho}
\def\s{\sigma}
\def\t{\tau}
\def\th{\theta}
\def\ii {\'\i  }

\title{\LARGE \bf  The energy states of  quantum dots 
through the variational method and SQM\footnote{Talk given at the VII International Wigner Symposium, 
University of Maryland, College Park, USA, August 2001}}
\author{Elso Drigo Filho}
\address{Instituto de Bioci\^encias, Letras e Ci\^encias Exatas,
IBILCE-UNESP\\ Rua Cristov\~ao Colombo, 2265 -  15054-000 S\~ao Jos\'e do Rio Preto - SP}
\author{Regina Maria  Ricotta}
\address{Faculdade de Tecnologia de S\~ao Paulo, FATEC/SP-CEETPS-UNESP  \\ Pra\c ca  Fernando
Prestes, 30 -  01124-060 S\~ao Paulo-SP\\ Brazil}
\maketitle

\begin{abstract}
The Variational Method is applied within the context of \-Supersymmetric Quantum Mechanics to
provide information about the energy states of a hydrogenic  donor in a quantum dot.
\end{abstract}

\section{Introduction}
Twenty years ago Supersymmetric Quantum Mechanics (SQM) appeared  inside the study of the SUSY
breaking mechanism of higher dimensional quantum field theories,
\cite{Witten}. Since then it has been considered as a new field of research, providing not
only a supersymmetric interpretation  of the Schr$\ddot {o}$dinger  equation, but interesting
answers to all sorts of non-relativistic quantum mechanical systems.  Particular points to be
mentioned include the better understanding it brought of  the exactly solvable,
\cite{Gedenshtein}-\cite{Levai},   the partially solvable,
\cite{Drigo1}, \cite{ Drigo2}, the isospectral, \cite {Drigo3}, \cite{Dunne}, the periodic,
\cite{Sukhatme} and the non-exactly solvable potentials, \cite{Drigo4}-\cite{Drigo6}. The latter
were studied by the association of the variational method with SQM formalism. The scheme employed
in them is based on  an {\it ansatz} made to the superpotential. Through the superalgebra  it is
possible to evaluate the  wave function, the trial wave function, which contains the free
parameters that minimize the energy expectation value.

We use this methodology to study the confined hydrogen atom, a system described by the confined
Coulomb potential. The confining approach is the same as the one  used to study the problem of a
hydrogenic donor located at the centre of a spherical $GaAs-(Ga,Al)As$ quantum dot, a
semiconductor device that confines electrons,  \cite{Zhu}, \cite{Varshni}.
Here the energy of the $1s$, $2p$ and $3d$ states of such confined system is evaluated and the
results  are compared to other variational and exactly numerical results, \cite{Varshni}.

\section{The variational method associated to SQM}
Consider a system described by a given potential $V_1$. The associated Hamiltonian $H_1$ can be
factorized in terms of  bosonic operators, in $\hbar = c = 1$ units, \cite{Sukumar}-\cite{Cooper1}.

\be
H_1 =  -{1\o 2}{d^2 \o d r^2} + V_1(r) =  A_1^+A_1^-  + E_0^{(1)}
\ee
where $ E_0^{(1)}$ is the lowest eigenvalue.  Notice that the function  $V_1(r)$ includes the
barrier potential term.  The bosonic operators are  defined in terms of the so called
superpotential $W_1(r)$, \be A_1^{\pm} =  {1\o \sqrt 2}\left(\mp {d \o dr} + W_1(r) \right) .
\ee As a consequence of the factorization of the Hamiltonian $H_1$,   the Riccati equation must
be satisfied,
\be
\label{Riccati} W_1^2 - W_1'=  2V_1(r) - 2E_0^{(1)}.
\ee Through the superalgebra, the eigenfunction for the lowest state is related to the
superpotential $W_1$ by
\be
\label{eigenfunction}
\Psi_0^{(1)} (r) = N exp( -\int_0^r W_1(\bar r) d\bar r).
\ee

At this point we address to the variational method. It was conceived to be an approximative
method to evaluate the energy spectra of a given Hamiltonian $H$ and, in particular, its ground
state. Its central point is the  search for an optimum  wave-function $\Psi(r)$ that depends of a
number of parameters. This is called the trial  wave-function.  The approach consists in varying
these parameters  in the expression for the expectation value of the energy
\be
\label{energy} E = {\int{\Psi^* H \Psi dr}\over  {\int{\mid \Psi \mid^2 dr}}}
\ee
until it reaches its minimum value.  This value is an upper limit of the
energy level pursued.

Thus, using physical arguments an  {\it ansatz} is made to the superpotential  and, through the
superalgebra, the wave function is evaluated. This is our trial wave function  which depends
on a set of the free variational parameters,  which were introduced by the  {\it ansatz} to the
superpotential.

Notice that since the potential is non-exactly solvable, the Hamiltonian is not
exactly factorizable, in other words, there is no  superpotential that satisfies the Riccati
equation exactly. However, the superpotential we have in hands through the {\it ansatz} does
satisfy the Riccati equation by an effective potential, $V_{eff}$
\be
\label{Veff}
V_{eff}(y) = {\bar W_1^2 - \bar W_1'\o 2}+ E(\bar\mu)
\ee
where $ \bar W_1 = W_1(\bar\mu)$ is the superpotential that satisfies (\ref{Riccati}) for
$\mu = \bar\mu$, the set of parameters that minimise the energy of eq.(\ref{energy}).

\section{The Confined Coulomb Potential}
The Hamiltonian of an on-center impurity in a spherical quantum dot can be written in the
effective-mass approximation as
\be H = -{\hbar^2 \o 2m^*} \nabla^2 - {e^2\o \epsilon r}
\ee where $m^*$ is the effective mass and $\epsilon$ is the dielectric constant of  the material
of the quantum dot. The donor is assumed to be at the centre of the quantum dot of  radius $R$
with an infinite barrier height. This means that the wave function vanishes at
$r=R$.

In atomic units, the radial Hamiltonian equation for the Coulomb Potential is given by
\be
\label{Coulomb} H = - {d^2 \o dr^2}  + {l(l+1) \o r^2}- {2\o  r}.
\ee We use the variational method associated with the SQM in order to get the energy states  of
the confined atom.

We make the following  {\it ansatz}  to the superpotential
\be
\label{W} W(r) =  - {c \o r} + b + {a \o R - r}
\ee which depends on three free parameters, the variational parameters. From the
superalgebra, the eigenfunction obtained from equation (\ref{eigenfunction}) is given by
\be
\Psi(a,b,c,r) \propto e^{-b r}\;  r^c \; (R - r)^ a.
\ee
This expression is a trial wavefunction for the variational method having three parameters, $a$,
$b$ and $c$ and vanishing at $r = R$.

The energy  is obtained by minimisation of the energy expectation value with respect to the
free  parameters.  The equation to be  minimised is given by the following evaluation
\be
\label{energymu} E(a,b,c) = {\int_0^{R-r} \Psi(a,b,c,r)  [- {d^2 \o dr^2} - {2\o r}  + {l(l+1)\o
r^2}]
\Psi(a,b,c,r) dr
\o \int_0^{R-r} \Psi(a,b,c,r) dr}.
\ee
In support of the choice made in (\ref{W}) for the superpotential are the physical arguments: the
first two terms of (\ref{W})  are related to the exact Coulomb case; the last term is the
confining term. This is remarked by the form of the effective potential $V_{eff}$ that satisfies
the Riccati equation, (eq. \ref{Veff}), evaluated at the values of $a$, $b$, and
$c$ that minimize the energy  expectation value,
\br V_{eff}(r,a,b,c) = {c(c-1) \o r^2} + {a(a-1)\o (R-r)^2} - {2bc \o r} - {2ac \o r(R-r)} + {2ab
\o R-r} + b^2 + \bar E(a, b, c) \nonumber
\er which is infinite at $r = R$.

\section{Results}
Thus, minimizing the equation (\ref {energymu}) with respect to the parameters $a$, $b$ and $c$
we find the minimum  value of the energy for different  values of $R$ and  $l$. These results
are shown in the tables bellow. Comparison is made with results from
\cite{Varshni}, which are based on the standard variational method in which a
variational wavefunction is proposed. Ref. \cite{Varshni} also contains exact numerical results.
\vskip .5cm

{\bf Table 1.}  Energy eigenvalues (in Rydbergs) for different values of $R$ for $l=0$ with a, b
and c as variational parameters. Comparison is made with results of Ref. \cite{Varshni}.\\
\begin{tabular}{|l|c|c|c|c|c|} \hline
\multicolumn{1}{|c} {R} &
\multicolumn{1}{|c|} {$E_{EXACT}$} &
\multicolumn{1}{|c|} {$E_V$} &
\multicolumn{1}{|c|} {$\underline{|E_{EXACT}-E_{V}|}\%$}  &
\multicolumn{1}{|c} {$E_{SQM}$} &
\multicolumn{1}{|c|} {$\underline{|E_{EXACT}-E_{SQM}|}\%$} \\
\multicolumn{1}{|c} {} &
\multicolumn{1}{|c} {Ref. \cite{Varshni}} &
\multicolumn{1}{|c} {Ref. \cite{Varshni}} &
\multicolumn{1}{|c} {$E_{EXACT}$} &
\multicolumn{1}{|c|} {  } &
\multicolumn{1}{|c|} {$E_{EXACT}$}  \\ \hline
0.1 & 937.986  & 937.999  & 0.00 & 940.688 & 0.29 \\ \hline
0.5 & 29.496 & 29.497 & 0.00 & 29.571 & 0.25 \\ \hline
1 & 4.7480  & 4.7484 & 0.01 & 4.7565 & 0.18 \\ \hline
2 & -0.25000  & -0.24990 & 0.04 & -0.25000 & 0.00 \\ \hline
3 & -0.84793  & -0.84523 & 0.32 & -0.84706 & 0.10 \\ \hline
4 & -0.96653  & -0.95518 & 1.17 & -0.96509 & 0.15 \\ \hline
\end{tabular}\\
\vskip .5cm

{\bf Table 2.}  Energy eigenvalues (in Rydbergs) for different values of $R$ for $l=1$ with a, b
and c as variational parameters. Comparison is made with results of Ref. \cite{Varshni}.\\
\begin{tabular}{|l|c|c|c|c|c|} \hline
\multicolumn{1}{|c} {R} &
\multicolumn{1}{|c|} {$E_{EXACT}$} &
\multicolumn{1}{|c|} {$E_V$} &
\multicolumn{1}{|c|} {$\underline{|E_{EXACT}-E_{V}|}\%$}  &
\multicolumn{1}{|c} {$E_{SQM}$} &
\multicolumn{1}{|c|} {$\underline{|E_{EXACT}-E_{SQM}|}\%$} \\
\multicolumn{1}{|c} {} &
\multicolumn{1}{|c} {Ref. \cite{Varshni}} &
\multicolumn{1}{|c} {Ref. \cite{Varshni}} &
\multicolumn{1}{|c} {$E_{EXACT}$} &
\multicolumn{1}{|c|} {  } &
\multicolumn{1}{|c|} {$E_{EXACT}$}  \\ \hline
 0.4 & 116.896  & 116.925 & 0.02 & 117.038 & 0.12 \\ \hline
2.0 & 3.1520  & 3.1530 & 0.03 & 3.1555 & 0.11 \\ \hline
4.0 & 0.28705   & 0.28706  & 0.00 & 0.28732  & 0.09  \\ \hline
6.0 & -0.111111  & -0.11069 & 0.38  & -0.111111 & 0.00 \\ \hline
7.0  & -0.17496   & -0.17392  & 0.59 &  -0.17490 & 0.03  \\ \hline
8.0 & -0.20890  & -0.20691 & 0.95 & -0.20882 & 0.04 \\ \hline
\end{tabular}\\
\vskip .5cm

{\bf Table 3.}  Energy eigenvalues (in Rydbergs) for different values of $R$ for $l=2$ with a, b
and c as variational parameters. Comparison is made with results of Ref. \cite{Varshni}.\\
\begin{tabular}{|l|c|c|c|c|c|} \hline
\multicolumn{1}{|c} {R} &
\multicolumn{1}{|c|} {$E_{EXACT}$} &
\multicolumn{1}{|c|} {$E_V$} &
\multicolumn{1}{|c|} {$\underline{|E_{EXACT}-E_{V}|}\%$}  &
\multicolumn{1}{|c} {$E_{SQM}$} &
\multicolumn{1}{|c|} {$\underline{|E_{EXACT}-E_{SQM}|}\%$} \\
\multicolumn{1}{|c} {} &
\multicolumn{1}{|c} {Ref. \cite{Varshni}} &
\multicolumn{1}{|c} {Ref. \cite{Varshni}} &
\multicolumn{1}{|c} {$E_{EXACT}$} &
\multicolumn{1}{|c|} {  } &
\multicolumn{1}{|c|} {$E_{EXACT}$}  \\ \hline
1.0 & 29.935  & 29.950 & 0.05 & 29.958 & 0.08 \\ \hline
3.0  &  2.5856  & 2.5867  & 0.04 &2.5876 &  0.08 \\ \hline
4.0  &  1.2427  &  1.2431 & 0.03 & 1.2437 &  0.08 \\ \hline
5.0 & 0.65823  & 0.65836 & 0.02 & 0.65873 & 0.08\\ \hline
7.0  &  0.19318  & 0.19318  & 0.00 &0.19333  & 0.08   \\ \hline
8.0 & 0.09212   & 0.09216  & 0.05 &  0.09220 & 0.08  \\ \hline
12.0 & -0.06250   &  -0.06181 &  1.11& -0.06250  & 0.00  \\ \hline
14.0  &   -0.08623 &  -0.08479& 1.66 & -0.08622 & 0.01  \\ \hline
\end{tabular}

\section{Comments and conclusions}
The variational method associated with SQM was applied to get the energy states of
a confined hydrogen atom. The problem is similar to the confinement of
electrons in a quantum dot. We used a confining effective potential that depends on tree
parameters (a, b, c). Minimizing the energy expectation value with respect to these three
parameters  we found good results, even for increasing values of the radius $R$, when
compared to results coming from other approximative variational method and exact numerical
results.
The term ${1\over (R-r)^2}$ in the effective potential leads to border effects, since it is
infinite at $R=r$. These effects do not appear in the original problem, eq. (\ref{Coulomb}) and
become more perceptive for  smaller values of $R$. For increasing values of $R$, smaller will be
the border effects so that the variational results get better.

In conclusion, we remark that the results presented here once again showed that the association
of  the superalgebra of SQM with the variational method provides good information about atomic
systems.
\section{Acknowledgements}
RMR and EDF would like to acknowledge partial financial support
from FAPESP and CNPq respectively.


\begin{thebibliography}{99}
\bibitem{Witten} E. Witten, Nucl.Phys. {\bf B188} 513 (1981)
\bibitem{Gedenshtein} L. Gedenshtein and I. V. Krive, So. Phys. Usp. {\bf 28} (1985) 645
\bibitem{Levai} G. L\'evai, Lect. Notes in Phys. {\bf 427} (1993) 427, Ed. H. V. von Gevamb,
Springer-Verlag
\bibitem{Drigo1} E. Drigo Filho, Mod. Phys. Lett. {\bf A9} (1994) 411
\bibitem{Drigo2} E. Drigo Filho and R. M. Ricotta, Physics of Atomic Nuclei {\bf 61} (1998) 1836
\bibitem{Drigo3} E. Drigo Filho, J. Math. Phys. {\bf A21} (1988) L1025
\bibitem{Dunne} G. Dunne and J. Feinberg, Phys. Rev. {\bf D57} (1998) 1271
\bibitem{Sukhatme} A. Khare and U. Sukhatme, J. Math. Phys. {\bf 40} (1999) 5473
\bibitem{Drigo4} E. Drigo Filho and R. M. Ricotta, Mod. Phys. Lett. {\bf A10} (1995) 1613
\bibitem{Drigo5} E. Drigo Filho and R. M. Ricotta, Phys. Lett. {\bf A269} (2000) 269
\bibitem{Drigo6} E. Drigo Filho and R. M. Ricotta, Mod. Phys. Lett. {\bf A15} (2000) 1253
\bibitem{Zhu} J. L. Zhu, J. J. Xiong, B. L. Gu, Phys. Rev. {\bf B 41} (1990) 6001; D. S. Chuu, C.
M. Hsiao, W. N. Mei, Phys. Rev. {\bf B 46} (1992) 3898;  N. Porras-Montenegro, S. T.
Perez-Merchancano, Phys. Rev. {\bf B 46} (1992) 9780; H. Parades-Gutierrez, J. C. Cuero-Yepez, N. Porras-Montenegro, J. Appl. Phys.
{\bf 75} (1994) 373
\bibitem{Varshni} Y. P. Varshni, Phys. Lett. {\bf 252A} (1999) 248
\bibitem{Sukumar} C. V. Sukumar, J. Phys. A: Math. Gen. {\bf 18} (1985) L57
\bibitem{Sukumar2} C. V. Sukumar, J. Phys. A: Math. Gen. {\bf 18} (1985) 2917
\bibitem{Cooper1} F. Cooper, A. Khare and U. P. Sukhatme, Phys. Rep. {\bf 251} (1995) 267

\end{thebibliography}
\end{document}